\documentstyle[twoside,fleqn,espcrc2,epsf]{article}

\newcommand{\geqsim}{\,\raisebox{-0.6ex}{$\buildrel > \over \sim$}\,}
\newcommand{\tr}{\mbox{Tr}}
\newcommand{\nn}{\nonumber}
\newcommand{\bea}{\begin{eqnarray}}   
\newcommand{\eea}{\end{eqnarray}}
\newcommand{\beq}{\begin{equation}}
\newcommand{\eeq}{\end{equation}}
\newcommand{\be}{\beta}

\title{Mass Spectrum of the 3d SU(2) Higgs Model and the
Symmetric Electroweak Phase\thanks{Talk presented by O. Philipsen}}

\author{O. Philipsen$^{\rm a}$,
M. Teper\address{Theoretical Physics, University of
Oxford, 1 Keble Road, Oxford OX1 3NP, UK} and 
H. Wittig\address{DESY-IfH Zeuthen,Platanenallee 6, D-15738 Zeuthen,
Germany}}

\begin{document}

\begin{abstract}
We present results for the masses of the low-lying states with 
quantum numbers $0^{++}$, $2^{++}$ and $1^{--}$ as well
as Polyakov line correlations in the Higgs
and confinement regions of the 3d SU(2) Higgs model.
In the confinement phase we find a dense spectrum of bound states  
approximately split into two disjoint sectors. One consists of W-balls 
nearly identical to the glueball spectrum of
the pure gauge theory, the other of bound states of scalars.
\end{abstract}

\maketitle

\section{INTRODUCTION}

It is by now well established that
the electroweak phase transition is of first order
for Higgs masses up to $m_H\sim 70\,{\rm GeV}$ \cite{ewpt}, and
there is strong evidence for a crossover behaviour for 
$m_H\geqsim 80\,{\rm GeV}$ \cite{kaj96}. 
While the problem of the order of the electroweak
phase transition thus seems to be solved, some puzzles remain
concerning the nature of the symmetric phase. In particular, computing
the mass spectrum of the 3d effective theory from gauge-invariant composite
operators leads to a picture of a confining symmetric phase with a 
spectrum of bound states \cite{ptw96}. 
However, calculations of the full propagators for the Higgs and
W bosons in Landau gauge
suggest significantly lighter states 
in the symmetric phase, while they yield similar masses in the Higgs 
phase \cite{bp94}. 

Here we elaborate on our gauge-invariant computation of the mass spectrum in
\cite{ptw96} by paying more attention to the excitation spectrum.
In addition to the $0^{++}$ and $1^{--}$ states considered previously,
we also compute the $2^{++}$ spectrum and the Polyakov line
correlations.
The three-dimensional lattice action is  
\bea \label{actlat} 
\lefteqn{S[U,\phi]=\be_G\sum_p\left(1-\frac{1}{2}\tr U_p\right)}
\nn \\
&&+\sum_x\Bigg\{-\be_H\sum_{\mu=1}^3\frac{1}{2}
\tr\Big(\phi^{\dagger}_x U_{\mu x}
\phi_{x+\hat{\mu}}\Big) \nn\\
&&+\frac{1}{2}\tr\Big(\phi^{\dagger}_x\phi_x\Big)
+\be_R\left[\frac{1}{2}\tr\Big(\phi^{\dagger}_x\phi_x\Big)-1\right]^2
\Bigg\}.
\eea
In this simulation we kept the ratio 
\beq
 \frac{\lambda_3}{g_3^2} = \frac{\beta_R\,\beta_G}{\beta_H^2} = 0.0239
\eeq
fixed and chose a point in the Higgs and confinement phase, 
$\be_H=0.3450$ and $\be_H=0.3438$, respectively, at $\be_G=9$.  

\section{IMPROVED OPERATORS AND CROSS CORRELATIONS}

We measure correlation functions of gauge-invariant operators of three
basic types. In the $0^{++}$ and $2^{++}$ channels we consider 
operators containing
only scalar fields, $R\sim \tr(\phi^{\dagger}_x\phi_x)$,  
scalar fields and links, 
$L_\mu\sim \tr(\phi^{\dagger}_x U_{\mu x} \phi_{x+\hat{\mu}})$,
or only plaquettes $P\sim \tr U_p$. 
In the $1^{--}$ channel there is only one operator type,
$V_\mu^a \sim \tr(\tau^a\phi^{\dagger}_x U_{\mu x}
\phi_{x+\hat{\mu}})$.

In order to improve the projection properties of our operators 
we employ ``smearing" or
``blocking" techniques \cite{tep87}. These consist of constructing
non-local, 
composite field variables $\phi^c_x$, $U^c_{\mu x}$ by covariantly connecting
them with their neighbours and then building the operators
out of these smeared fields.
This procedure can be iterated to
create operators of various spatial extensions.
These are more sensitive to infrared physics and 
have a better overlap with states that represent extended
particles, such as weakly bound states. 

In order to compute the excitation spectrum of states with a given
quantum number we keep $N$ operators of different types and blocking 
levels and measure cross correlations between all of them.
The correlation matrix can then be diagonalised numerically following a
variational method. For a given set of operators ${\phi_i}$ we 
find the linear combination that minimises the energy
and thus corresponds to the lightest state. The first excitation
can be found by repeating this step, but restricted
to the subspace 
$\{\phi_i\}^{'}$ that is orthogonal to the ground
state. This procedure can be continued to higher states, and we end up
with a set of $N$ eigenstates,
%
$\Phi_i =  \sum_{k=1}^N a_{ik}\phi_k$.
%
The coefficients $a_{ik}$ 
are useful in identifying the 
contributions of the individual operators used in the simulation to the
mass eigenstates.

A detailed discussion of the construction of operators, the blocking
techniques adopted 
and the diagonalisation procedure
may be found in \cite{ptw96}. 

\section{THE MASS SPECTRUM}

\begin{figure}[hb]
\vspace{-1.1cm}
\begin{center}
\leavevmode
\epsfysize=200pt
\epsfbox[20 30 620 730]{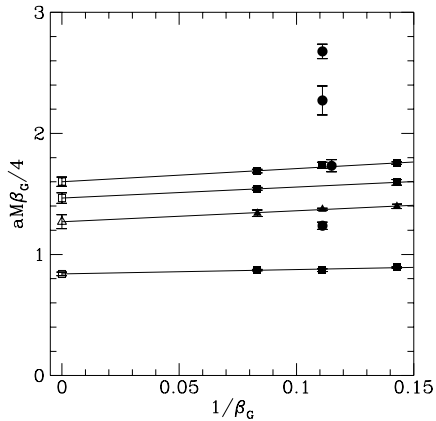}
\vspace{-2.5cm}
\end{center}
\caption[]{\label{conts}
 Continuum limit in the confinement region.
 Squares and circles represent the three lowest
 $0^{++}$ and $2^{++}$ states, triangles denote the lowest $1^{--}$
state.
 Open symbols indicate the data extrapolated to $1/\beta_G=0$.}
\vspace*{-0.5cm}
\end{figure}

\begin{figure}[ht]
\vspace{-0.4cm}
\begin{center}
\leavevmode
\epsfysize=200pt
\epsfbox[20 30 620 730]{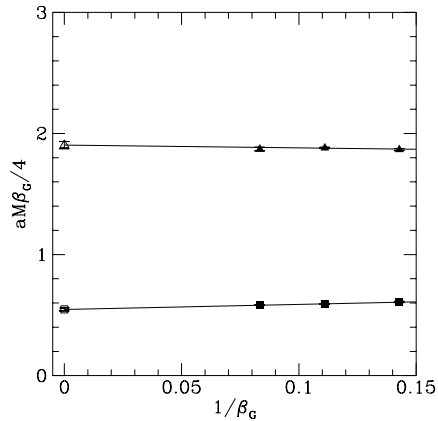}
\vspace{-2.5cm}
\end{center}
\caption[]{ \label{contb}
Continuum limit in the Higgs region.}
\vspace*{-0.5cm}
\end{figure}

The continuum extrapolation of the lowest 
$0^{++}$ and $1^{--}$ mass eigenstates 
of the spectrum as obtained in \cite{ptw96} 
is shown in Figs.~\ref{conts},\ref{contb}. 

\begin{figure}[th]
\vspace{-1.0cm}
\begin{center}
\leavevmode
\epsfysize=200pt
\epsfbox[20 30 620 730]{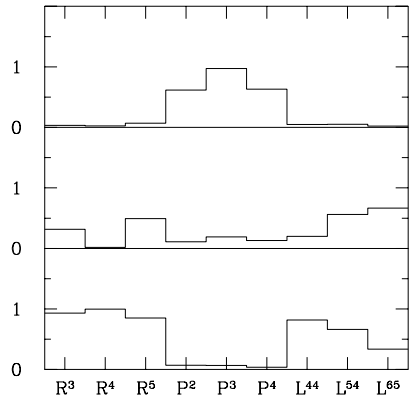}
\vspace{-2.5cm}
\end{center}
\caption[]{ \label{cross}
 The coefficients $a_{ik}$ of the 
$0^{++}$ operators used in the simulation. Indices at $R$,$P$,$L$
indicate different blocking levels, see \cite{ptw96}.} 
\vspace*{-0.5cm}
\end{figure}

In the Higgs phase higher excitations are heavy (the lightest being about twice
the mass of the W) and plateaux in the 
effective masses are hard to identify. This is not surprising since they
are expected to be scattering states with relative momentum.
In the symmetric phase, on the other hand, there is a dense spectrum of
bound states.
Their composition may be characterised by considering the contributions of 
the individual operators to each eigenstate, 
as shown in Fig.~\ref{cross} for the $0^{++}$ channel at $\be_G=12$.
The pure gauge degrees of freedom, $P$, 
contribute very little to the
ground state and the first excited state.
However, the third $0^{++}$ state is composed almost entirely of them.
This suggests interpreting it as a 
``W-ball", in analogy to the glueballs of pure gauge theory. 
Fig.~\ref{conts} also shows preliminary results for 
a set of $2^{++}$ operators. In this channel the lowest state is
composed predominantly of $R$ and $L$ type operators, the following two
excited states are mixed, and the third excited state receives $P$
contributions only, thus being interpreted as a $2^{++}$ W-ball.
In Table~1 the masses of the W-balls and their excitations 
are compared with those of the glueballs in the pure gauge theory, from
which they differ at the percent level at most. 
\begin{table}[ht]
\vspace*{-0.5cm}
\label{comp}
\caption[]{ 
Purely gluonic excitations in the gauge
Higgs and pure gauge systems at $\beta_G=9$.}
\begin{tabular}{|l|r@{.}l|r@{.}l|}
\hline
  & \multicolumn{2}{c|}{gauge Higgs}
  & \multicolumn{2}{c|}{pure gauge} \\ \hline
$aM[0^+]$ & 0&751(8) & 0&767(6) \\ \hline
$aM^*[0^+]$ & 1&07(2) & 1&08(2)\\ \hline
$aM^{**}[0^+]$ & 1&26(3) & 1&27(2)\\ \hline
$aM[2^+]$ & 1&19(3) & 1&26(2) \\ \hline
$aM^*[2^+]$ & 1&38(4) & 1&50(5) \\ \hline
$aM^{**}[2^+]$ & 1&76(10) & 1&77(6) \\ \hline
$a\sqrt{\sigma}$ & 0&1582(6)& 0&1616(6)\\
\hline
\end{tabular}
\vspace*{-0.5cm}
\end{table}

\section{POLYAKOV LOOPS}

Another quantity that may help to clarify the nature of the
symmetric phase is the Polyakov line operator whose
expectation value vanishes in a confining theory.
One would not expect it to be exactly zero in the
symmetric phase of the Higgs model because flux tubes
between fundamental charges eventually break with increasing
separation. At our parameter values in the symmetric
phase, however, the VEV of the Polyakov line operator is
zero within errors. Hence we may extract a volume-corrected string tension
from Polyakov line correlations according to \cite{for85},
\beq
a^2\sigma=a^2\sigma_L+\frac{\pi}{6}\frac{1}{L^2};\quad
aM_{Pol}(L)=a^2\sigma_LL\; ,
\eeq 
where $L$ is the side length of the lattice.
In the symmetric phase we find $aM_{Pol}(24)=0.579(4)$.
This yields a string tension $\sim 97\%$ of the one found in the pure gauge
theory, c.f.~Table~1.
In the Higgs phase we estimate $aM_{Pol}=1.8(1)$.
Here the Polyakov loop operator exhibits a VEV, no flux tubes exist, 
and the correlator cannot be
related to a string tension. Instead, a weak coupling expansion is
possible whose leading term is a two-W exchange diagram.
Indeed, the effective mass for the Polyakov line is consistent with 
twice the W-mass, $aM_W=0.836(2)$ at $\be_G=9$.

\section{CONCLUSIONS}

To summarise, we have confirmed the approximate decoupling of the pure
gauge sector from the Higgs sector in the symmetric phase.
The mass spectrum includes a set of W-balls, which is essentially equivalent
to the glueball spectrum in pure gauge theory and undisturbed by the
presence of matter fields.
In addition there are
also bound states of scalars, seemingly rather disconnected from the
W-ball part of the spectrum. The effective vanishing of the Polyakov loop
VEV as well as the value of the string tension are further
similarities of the symmetric phase with the pure gauge theory. 
More work is needed to understand the relation between the
gauge-fixed and gauge-invariant calculations in the symmetric phase.
It is interesting to speculate whether the same decoupling of
glueballs holds in QCD.

\end{document}